%% file: prepost.tex
\documentclass[copyright,creativecommons]{eptcs}

\providecommand{\event}{PrePost 2016} 
\usepackage{breakurl}             
\usepackage[T1]{fontenc}
\usepackage{url}
\usepackage{listings}
\usepackage{graphicx}
\usepackage{caption}
\usepackage{subcaption}
\usepackage{calc}  
\usepackage{enumitem}
\usepackage{amsthm}
\usepackage{lipsum}
\usepackage{wrapfig}
\usepackage{dirtree}

\DeclareFixedFont{\ttb}{T1}{txtt}{bx}{n}{10} 
\DeclareFixedFont{\ttm}{T1}{txtt}{m}{n}{10}  

\usepackage{color}
\definecolor{deepblue}{rgb}{0,0,0.5}
\definecolor{deepred}{rgb}{0.6,0,0}
\definecolor{deepgreen}{rgb}{0,0.5,0}

\newcommand\pythonstyle{
    \lstset{
        language=Python,
        basicstyle=\ttm,
        otherkeywords={self},             
        keywordstyle=\ttb\color{deepblue},
        emph={MyClass,__init__},          
        emphstyle=\ttb\color{deepred},    
        stringstyle=\color{deepgreen},
        frame=tb,                         
        showstringspaces=false            %
    }
}

\DeclareCaptionFormat{listing}{\rule{\dimexpr\textwidth\relax}{0.3pt}\par\vskip1pt#1#2#3}
\newcounter{codecounter}

\lstnewenvironment{python}[1][]
{
\refstepcounter{codecounter}
\pythonstyle
\lstset{#1}
}
{}

\title{Model-based Testing of Mobile Systems -- An Empirical Study on QuizUp Android App}

\author{Vignir Gudmundsson
\institute{Reykjavik University\\
Reykjavik, Iceland}
\email{vignirgudmunds@gmail.com}
\and
Mikael Lindvall
\institute{Fraunhofer CESE\\
Maryland, USA}
\email{mlindvall@fc-md.umd.edu}
\and
Luca Aceto
\institute{Reykjavik University\\
Reykjavik, Iceland}
\email{luca@ru.is}
\and
Johann Bergthorsson
\institute{Plain Vanilla Games\\
Reykjavik, Iceland}
\email{johann@plainvanillagames.com}
\and
Dharmalingam Ganesan
\institute{Fraunhofer CESE\\
Maryland, USA}
\email{dganesan@fc-md.umd.edu}
}
\def\titlerunning{Model-based Testing of Mobile Systems}
\def\authorrunning{Gudmundsson et al.}

\begin{document}

\newcounter{examplecounter}
\newenvironment{example}{\begin{quote}%
    \refstepcounter{examplecounter}%
  \textbf{Example \arabic{examplecounter}}%
  \quad
}{%
\end{quote}%
}

\newcounter{issuecounter}
\newenvironment{mww_issue}{\begin{quote}%
  \textbf{Sample issue related to middleware wrappers}%
  \quad
}{%
\end{quote}%
}

\newenvironment{api_issue}{\begin{quote}%
  \textbf{Sample issue related to GMSEC Java API}%
  \quad
}{%
\end{quote}%
}

\newenvironment{ui_issue}{\begin{quote}%
  \textbf{Sample issue related to updating user information}%
  \quad
}{%
\end{quote}%
}

\newcommand{\todo}[1]{
  \vspace{-2mm}
  \begin{center}
  \colorbox{cyan}{
    \begin{minipage}{0.95\linewidth} \footnotesize
      \textbf{Todo:} #1
     \end{minipage}
     }
 \end{center}
}

\newcommand{\vignir}[1]{
  \vspace{-2mm}
  \begin{center}
  \colorbox{cyan}{
    \begin{minipage}{0.95\linewidth} \footnotesize
      \textbf{Vignir:} #1
     \end{minipage}
     }
 \end{center}
}

\newcommand{\mikael}[1]{
  \vspace{-2mm}
  \begin{center}
  \colorbox{cyan}{
    \begin{minipage}{0.95\linewidth} \footnotesize
      \textbf{Mikael:} #1
     \end{minipage}
     }
 \end{center}
}

\newcommand{\luca}[1]{
  \vspace{-2mm}
  \begin{center}
  \colorbox{cyan}{
    \begin{minipage}{0.95\linewidth} \footnotesize
      \textbf{Luca:} #1
     \end{minipage}
     }
 \end{center}
}

\maketitle

\input{chapters/abstract}
\input{chapters/01_introduction}
\input{chapters/02_mbt_approach}
\input{chapters/03_quizup_study}

\input{chapters/04_discussions}
\input{chapters/05_conclusions}

\paragraph*{Acknowledgements:}
\label{sec:acknowledgements}
This research was partly supported by the project ``TheoFoMon:
Theoretical Foundations for Monitorability'' (grant number:
163406-051) of the Icelandic Research Fund.


\end{document}

%% file: chapters/abstract.tex
\begin{abstract}
We present an empirical study in which model-based testing (MBT) was applied to a mobile system: the Android client of QuizUp, the largest mobile trivia game in the world. The study shows that traditional MBT approaches based on extended finite-state machines can be  used to test a mobile app  in an effective and efficient way. Non-trivial defects were detected on a deployed system that has millions of users and was already well tested. The duration of the overall testing effort was of three months, including the construction of the models. Maintaining a single behavioral model for the app was key in order to test it in an efficient way. 
\end{abstract}

%% file: chapters/01_introduction.tex
\section{Introduction}\label{chap:introduction}

Testing of mobile systems (apps) is a new challenge for software companies. Mobile apps are often available on multiple platforms, operating systems, programming languages, etc. This makes it necessary to ensure that the app behaves in the same way independently of the platform on which it runs and of what language is used. In addition, mobile systems often have many configuration options that influence the way they behave. Thus app developers must ensure that the app works as expected for different configurations. Apps are often tested manually. Such manual testing typically relies on the manual creation and execution of test cases that mimic realistic usage of the app. Manual testing is tedious, and is likely to be exhausting rather than exhaustive, especially when a large number of combinations of usage scenarios for various configurations must be covered. For each configuration, the tester has to manually enter data, manually swipe the screen, click on buttons, and manually compare the actual result and behavior with the expected one. Differences between actual and expected result and behavior are then manually documented and reported as issues. It is also very tedious and challenging for humans to validate the expected results of a graphical user interface (GUI) properly, because the state tends to be very verbose. For example, for every screen or state in an app, there are usually many elements to validate. This state of affairs should be compared with API testing, where one has simple error codes as responses.

Whenever a new version of the system under test (SUT) has been developed, the manual tester must go through the same process again. Since new versions are typically released several times per year,  and  sometimes even twice or thrice per month, the manual testing effort can be significant and is likely to miss key errors because of the large amount of elements to validate.

The problems related to manual testing have been observed by many software organizations who  strive to automate the testing process. This has led to the successful use of test case execution frameworks (e.g., JUnit~\cite{junitbook}). Such frameworks are very helpful because they can automatically run the tests  once they have been encoded as executable test cases (a.k.a. concrete test cases and test scripts/programs). This type of automation frees up the manual tester because the executable test case automatically enters data, issues commands, compares actual and expected values, and reports detected issues. JUnit is now available also for mobile apps.

However, such test programs must still be created and maintained manually, which can be difficult. One reason is that mobile apps typically have a significant number of screens, input fields and buttons and can be highly configurable. Another reason is that it can be hard to imagine the many different ways in which the app can be used or misused. Thus, the manual creation of executable test cases can be a very complex activity. 

In previous work~\cite{DBLP:conf/issre/GudmundssonSGLW13,DBLP:journals/isse/GudmundssonSGLW15,DBLP:conf/icse/LindvallGAW15}, we have used model-based testing (MBT) to test non-mobile systems because MBT addresses the problem of manually creating test cases. Instead of creating one test case at a time, MBT builds a model of the system under test, which is then used to automatically generate executable test cases. While MBT is a promising technology that has been shown to work well for non-mobile systems, it is still unclear if it works well also for mobile apps. 
This leads to the research question we address in this paper: \begin{quote}{\em Can MBT be used, in an effective and efficient way, to test mobile systems using the same approach that has been used to test non-mobile systems?}\end{quote}
The reason this is a valid question is that, even though there are many commonalities between a mobile app and a non-mobile app such as the types of controls (buttons, input fields etc.), there are also some differences, such as gestures in the form of swipes, and it was unclear if the modeling notation would allow us to describe these new input methods. In addition, MBT only works well when the app is running in a robust testing environment and it was not clear to us if the available testing environments would be suitable for the type of testing `our' version of MBT needed (e.g. request and response). It was also unclear if the app simulators would be powerful enough, and if not, if testing on a mobile device itself would be an option.
To answer our research question, we examined the mobile system QuizUp, which is a trivia game that supports both the Android and iOS mobile operating systems~\footnote{https://quizup.com/}. At the time of the study, QuizUp had over $400$ topics including over $220,000$ questions, which have grown to over $1,200$ topics and more than $600,000$ questions since then. We consider QuizUp a strong representative of a standard mobile system, i.e. a system that uses networking, stores data in databases, requires user authentication, contains most basic types of UI elements, etc.

We also address whether MBT can be applied efficiently to mobile apps. In the context of this research, efficient means `with reasonable effort' such as an internship where the duration is limited to six months, during which the person has to understand the system under test, build all testing infrastructure, build the model, generate test  cases, test the system, and analyze results. Our tenet is that if all of these activities could be achieved in such limited time, then it would be reasonable to conclude that the technology is applicable to a wide range of systems since we applied the same type of MBT to other types of systems in the past \cite{DBLP:conf/issre/GudmundssonSGLW13, DBLP:conf/icse/SchulzeGLCG14, DBLP:conf/issre/SchulzeGLOC13}. Thus, another goal of this work was to study the overall costs and benefits of using MBT in general. 

The results of our study show that MBT {\em is} feasible also for mobile apps because it was applied with reasonable effort and it detected several software bugs that, despite extensive effort, were not found by the traditional testing approach used at QuizUp. Summarizing, this paper contributes an industrial case study, with an analysis of the costs and benefits resulting from using extended finite state machine (EFSM) MBT on a mobile app.


\subsubsection{Structure of the paper}
In Section 2, we introduce the MBT approach, which is based on state-machines, that was used. In Section 3, we present the case study where MBT was applied to QuizUp's Android client. In Section 4, we compare and discuss the similarities and differences between applying MBT on QuizUp and on other systems, especially GMSEC. In that section, we also discuss some related work as well as some ideas for future work. In Section 5, we present and discuss our conclusions.




%% file: chapters/02_mbt_approach.tex
\section{Model-based testing with state machines}

\subsection{Model representations}\label{sec:model_repr}
We chose to use the EFSM style of modeling representations because state machines are easily understood and well-studied theoretically~\cite{sipser2006introduction}. Moreover, easy-to-use open source tools that generate test cases from them are readily available~\cite{jumbl}\footnote{http://www.graphwalker.com/}.

Informally, an EFSM consists of states and transitions~\cite{sipser2006introduction}. It contains a specific state, called the start state, where the machine starts computing. In the context of MBT, a transition represents a system stimulus, or action, that moves the user from one state of the system to another. Such actions could, for instance, be calling a method or clicking a button. The states are used to embed  assertions to check that the user is in the expected system state. These assertions are boolean expressions based on the return code that is expected from a particular method call when a system is in a certain state.


The EFSM model is a generalization of the traditional Finite-State Machine model with guards or helper functions associated with transitions~\cite{kita1999method}. Guards are boolean expressions or functions that are evaluated during model-traversal time. State variables allow one to store the history of traversal and can be used to embed test oracles in the model. Based on the evaluation of a guard, a model-traversing tool or algorithm chooses among the transitions whose guards are satisfied. Using EFSMs we are able to encode data parameters as variables as opposed to embedding the data at the model level as transitions or states. Such encoding can, for example, be done to parameterize, or initialize, the model at the start of each traversal, storing data as variables or other data structures. This allows one to modify the data to be tested without changing the model and thus it does not affect the number of states and transitions in the model using EFSMs. Hence, the model is easy to review, comprehend, and evolve. In addition, guards and helper functions can be used to return data from actions based on the current traversal history.


\subsection{Overview of MBT using state machines}\label{sec:MBToverview}
MBT uses a model to represent the expected behavior of the SUT, which is used to generate test cases.  Each test case is a path through the model.  Such a path can, for instance, be randomly generated---thus building a certain amount of randomness automatically in the testing process. MBT addresses the problem with test cases being static and not covering enough corner cases, which are often exercised by taking an unusual path through the system or using some functionality repeatedly and in unexpected ways.

The model is created from the perspective of a user focusing on the functionality of the SUT's \emph{interface}, where a user can be either a human being or another program/system. 
Thus, testing using this approach is driven by the specifications of the interface of the SUT and does not use the internal structure of the software under test as the driver~\cite{Tretmans2003}. The assumed benefit is that, even though the complexity of the SUT is large, the model and sub-models typically remain manageable in size. Thus, instead of depending on the complexity of the actual system, the models are as complex as the model of the interface under test.

The process is as follows:
1) A model is built by the tester based on the requirements, existing test cases, and by exploring the SUT. 
2) A tool traverses the model and generates  \emph{abstract test cases}, which are sequences of state and transition names.  
3) All state and transition names in the model are extracted programmatically and turned into a table that lists each state and transition name. This table constitutes the \emph{mapping table} when executable code fragments are manually entered by the tester for each state and transition.
4) An \emph{instantiator} program automatically creates executable test cases by replacing each name in the abstract test case with the corresponding code from the mapping table.  
5) The test cases are executed automatically. 
6) The failed test cases are analyzed by the tester. 
    ~ 

    ~ 

%% file: chapters/03_quizup_study.tex
\section{A Mobile Case Study: QuizUp}

QuizUp is a mobile trivia game that allows users to challenge each other on several hundred topics (e.g. arts, science, sports) using almost a quarter million questions at the time of this study. 
Users participate in a social experience by communicating and competing against friends or strangers in a real-time trivia quiz. The application, initially released for the iOS platform, turned out to be an overnight success\footnote{https://itunes.apple.com/us/app/quizup/id718421443}. Within a year since its initial release, QuizUp released an Android client of the application~\footnote{https://play.google.com/store/apps/details?id=com.quizup.core}, resulting in over $20$ million Android and iOS users.

Testing the application is a complex task due to its data-driven design, its complexity and the configuration options it provides. The data-driven design is embodied by the real-time dependence on large amounts of data to display. The application communicates with the QuizUp servers, through web services, to fetch data (e.g. HTTP GET query) from the QuizUp databases, as well as posting new or updated data to the databases (e.g. HTTP POST query). For most of the scenes in the application, a significant portion of the data being displayed depends on data from the QuizUp databases. Thus, as testers, we can only control data, or the order of data, beforehand in limited scenarios, as we do not have any authority over the databases.

The complexity of the app is largely due to the game-play scenes of the game. A user can compete against other users (a.k.a. opponents) in any of QuizUp's topics. After the user has requested to play a game on a particular topic, the system searches, in real-time, for an opponent who has also requested to play that particular topic at that particular moment. 


It is critical that the app runs on different multiple mobile platforms. The Android and iOS clients are native clients implemented by separate teams. The Android client is implemented in Java; the iOS client is implemented in Objective-C. Nevertheless, the clients should follow the same set of business rules and requirements, and therefore behave conceptually in the same way. 


Due to the above-mentioned testing challenges, the QuizUp team conducts large amounts of testing. The team has addressed ambitious testing goals, resulting in the rapid and astounding success of the application, but with a high cost and a substantial effort in thorough Quality Assurance (QA). 
The testing effort can be divided into several categories. The QuizUp team has developed a set of executable test cases, written in Calabash\footnote{https://github.com/calabash/calabash-ios}, that address some of the testing goals. However, the test suite is manually written and limited to short and common scenarios for the iOS client, and not for the Android client. The team had
no automated tests for the Android client; instead it managed a beta group of Android users that provided rapid feedback prior to new releases. The team also included five QA members who constantly verified that new versions and updates of the application met its business rules and requirements through manual regression testing. The QuizUp team had developed an extensive Excel sheet that outlined hundreds of singular tests (rows of action/expected outcome) that QA testers and third party testers used as reference. The team found it tedious and difficult to maintain this spreadsheet.
The team also outsourced a large set of end-user acceptance tests to a contractor company that assisted them with QA. Thus, the overall testing effort was significant.

Since the QuizUp team is interested in improving their testing processes through automation, our primary goal was to study the feasibility of using MBT on the QuizUp application.
This interest was sparked by the possibility that the QuizUp team would maintain the MBT models and infrastructure after the study. However, the learning curve for applying MBT on a mobile application, such as QuizUp, was unclear. Thus, another goal of the study was to clarify the needed effort.


Following the initial research question, we derived the following sub-questions: 1) Can the QuizUp application be modeled in such way that test cases can be automatically generated to test its core features?
2) Can we design the model in such way that it can test both the Android and iOS client without modifying the model? Since the two clients should represent the `same' QuizUp application, it was desirable to maintain a single QuizUp model instead of maintaining two separate models. Since the QuizUp team has thorough testing processes, we would consider the MBT approach to be successful if it were able to detect non-trivial issues. We decided to test the Android client of QuizUp through its GUI. Although we did not test the iOS client, the implemented testing approach was designed in such way that the iOS client could be easily integrated in future work. We chose the Android client over the iOS client because it was more accessible for test automation due to Apple's hardware restrictions\footnote{Apple requires automated tests for iOS applications to run on Mac OS X systems to which we had no access.}, and due to the fact that no automated tests for the Android client existed. The derived tests were run on mobile emulators using the Genymotion Android emulator\footnote{https://www.genymotion.com}. Although current emulator technology supports the emulation of the physical state of mobile devices, such as network reception, sensors, battery status, etc., we did not test for those activities and events. The study was mainly performed at QuizUp's facilities. Questions were asked to the QuizUp team during the process when the documentation was ambiguous or not specific enough. Apart from that, the effort was carried out independently of the QuizUp team. The findings were reported to the QuizUp team during and after the study. In particular, frequent interaction with the QuizUp team was crucial during the construction of the models because we did not have a concrete documentation or specification of the QuizUp system. 

\subsection{Core features of the QuizUp application}

The application is divided into \emph{scenes}, which are either \emph{access} scenes or \emph{in-game} scenes. Access scenes identify the user, through log-in or sign-up, prior to using the in-game scenes of the application. In-game scenes enable the logged-in user to use various features of the game, such as playing trivia question rounds against other users, communicating with opponents and viewing leader boards. Most scenes in the QuizUp application contain sub-scenes as well as sub-sub-scenes. Below, we will describe a few selected sub-scenes. 


    ~ 
    ~ 

\begin{figure}[t!]
    \centering
    \begin{subfigure}[b]{0.245\textwidth}
            \includegraphics[width=1.33in]{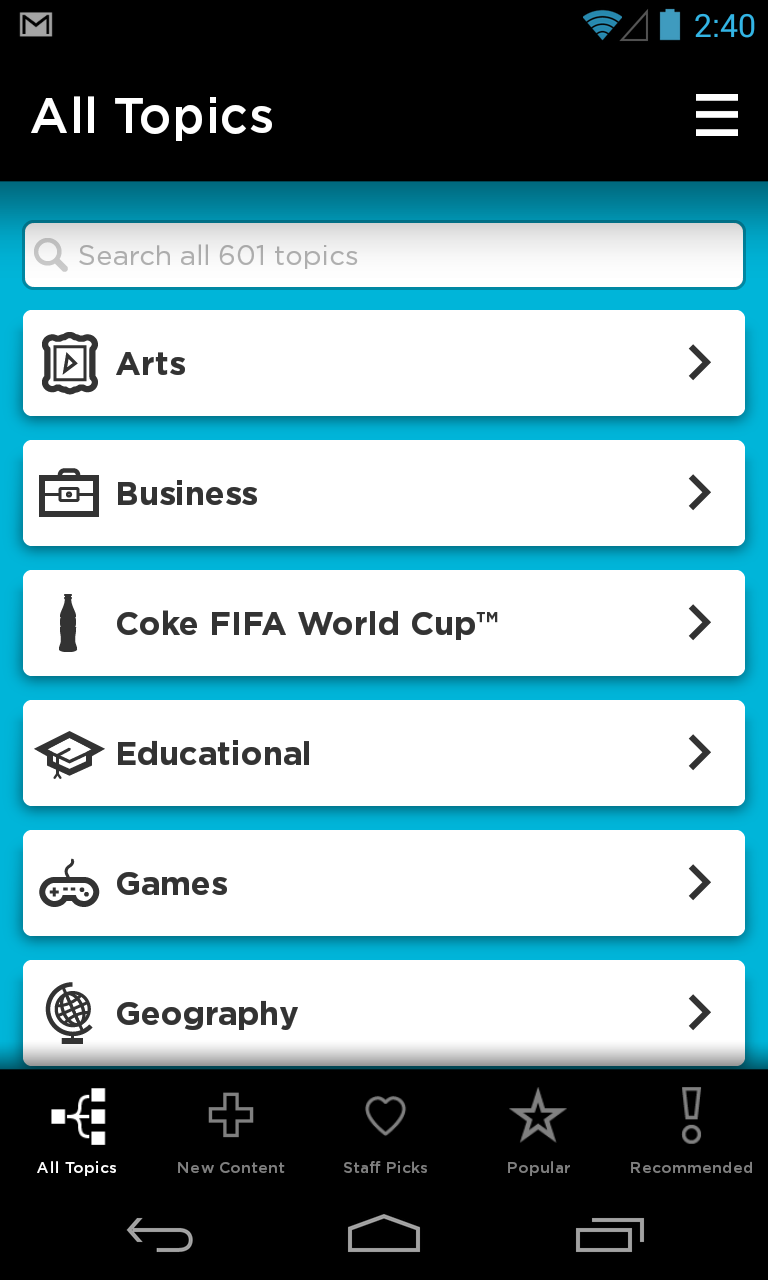}
            \caption{Topics scene}
            \label{fig:topics_scene}
    \end{subfigure}
    \begin{subfigure}[b]{0.245\textwidth}
            \includegraphics[width=1.33in]{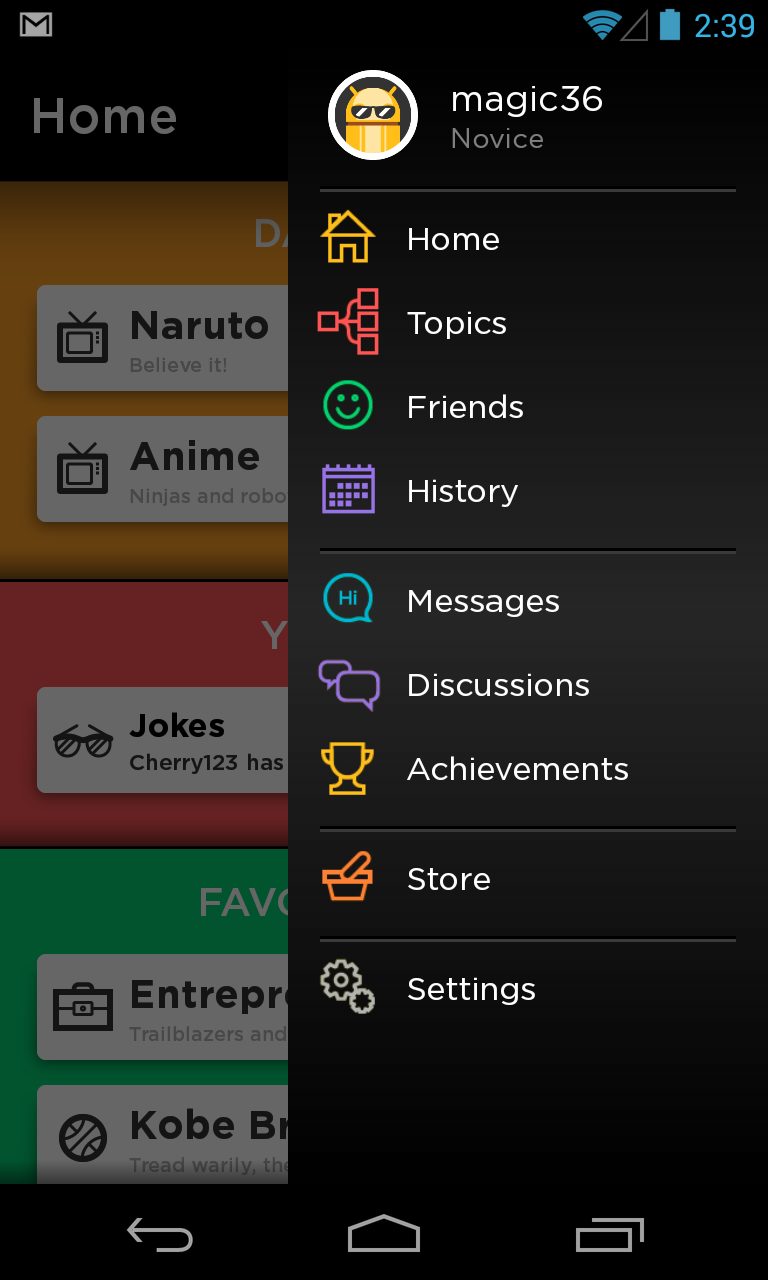}
            \caption{Game sidebar}
            \label{fig:sidebar}
    \end{subfigure}
    \begin{subfigure}[b]{0.245\textwidth}
            \includegraphics[width=1.33in]{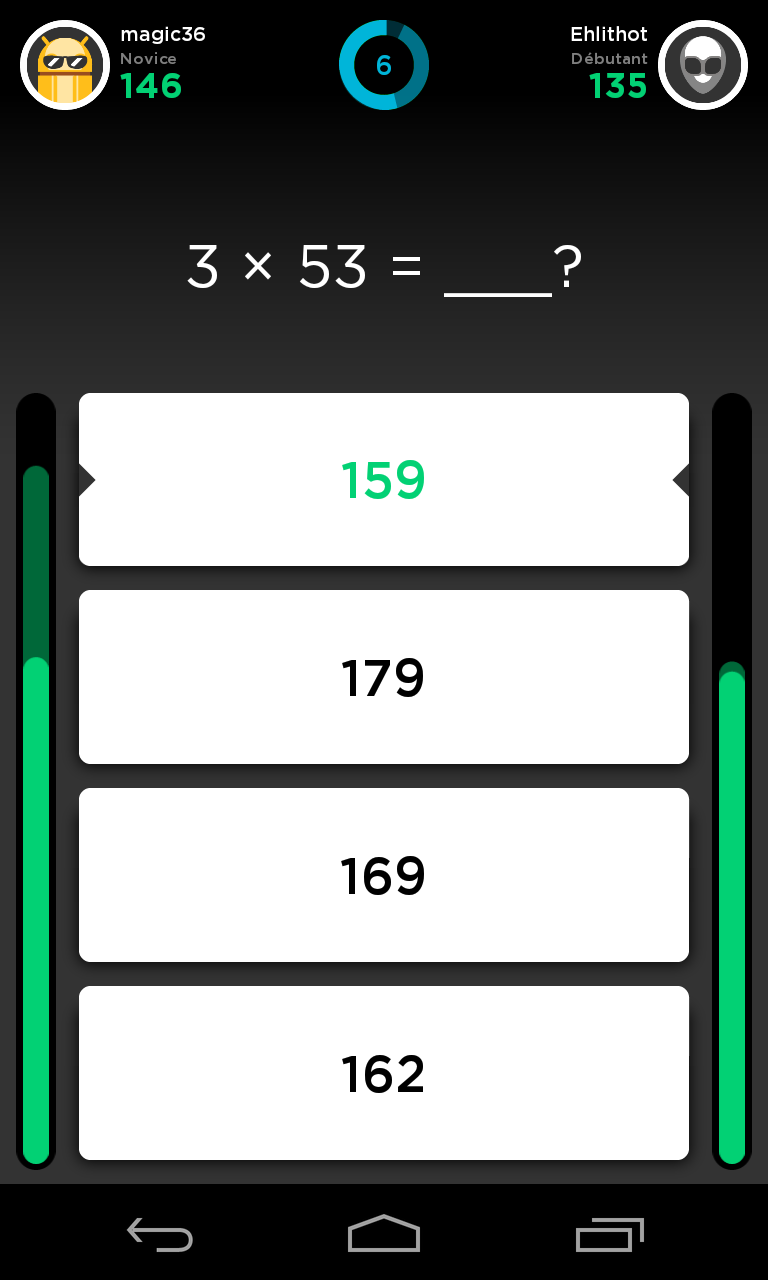}
            \caption{Game-play scene}
            \label{fig:gameplay6}
    \end{subfigure}
    \begin{subfigure}[b]{0.245\textwidth}
            \includegraphics[width=1.33in]{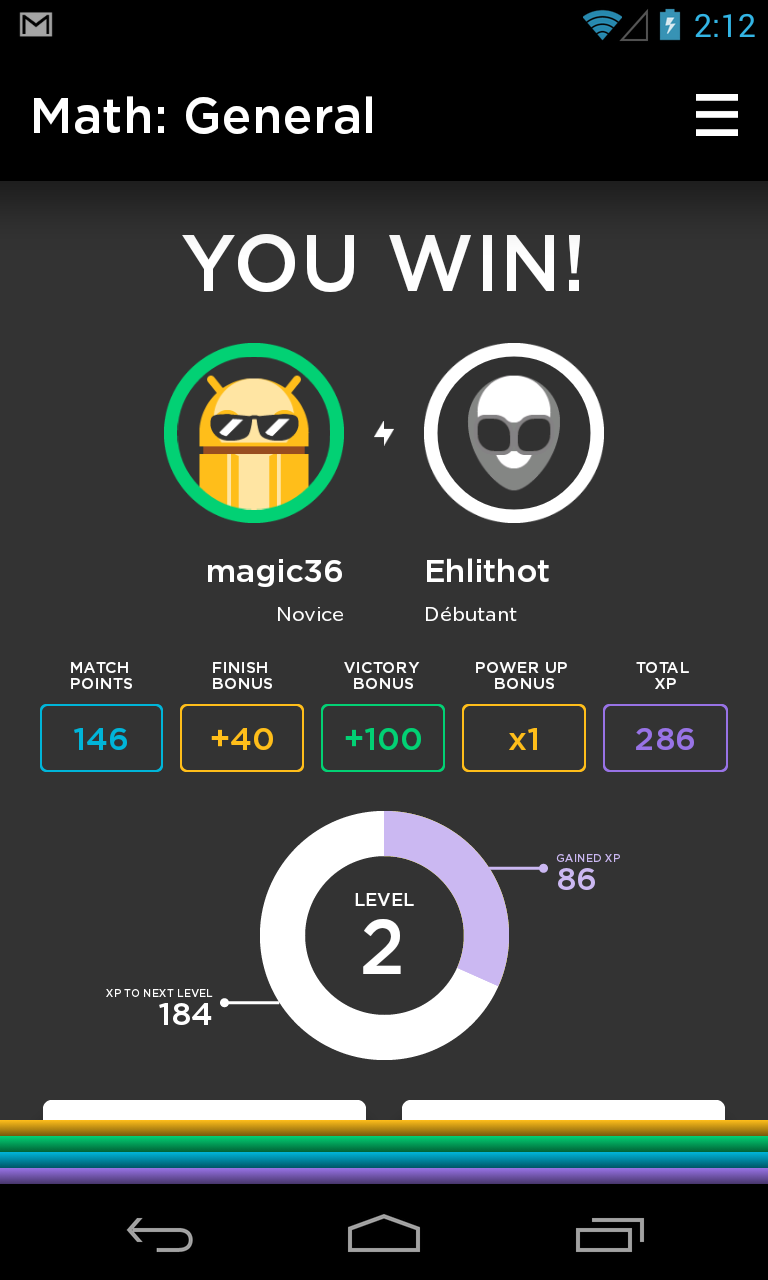}
            \caption{Game-stats scene}
            \label{fig:gameplay8}
    \end{subfigure}
    \caption{Several of QuizUp's in-game scenes.}\label{fig:in_game_scenes3}
\end{figure}

\subsubsection{Rules of QuizUp}
There are additional behaviors related to the sequence of actions and the navigation between scenes that we describe as \emph{rules} because they influence the  testing that needs to be performed.
1) First-time log-in: New users should see introductory information displayed in the Messages and History scenes.
2) Clean up upon logging out: When a user logs out through the Settings scene, the session data should be cleaned up. 
3) Profile changes should be recognized in all relevant scenes: Changing profile information, such as the user's name or title, should enforce an update in all other scenes that display that particular data.
4) Games should be stored in the History scene: After playing a game against an opponent, a record of that game should be stored in the History scene, where the user can view information and statistics about that game.
5) Sent messages in Chat conversation should be stored in the Messages scene: A message history with a particular opponent can be accessed through multiple different scenes in the application. The conversation should always be up to date independently of where it was entered.
6) The scene prior to entering the Settings scene matters. The Settings scene has two tabs, a settings tab and a profile tab, and which tab should be set depends on the scene prior to entering the Settings scene. The profile tab should be set when entering the Settings scene by using the shortcut button in the Profile scene. The settings tab should be set upon entering the Settings scene from any scene other than the Profile scene and the Settings scene itself. Attempts to enter the Settings scene while the Settings scene is open should be ignored.


\subsubsection{Testing questions}
From the scenes  and rules we derive the testing questions listed below. Some are specific to QuizUp while others also apply to other mobile applications.
Q1: Are there issues related to the access scenes displaying error messages when invalid information is input?
Q2: Are there issues related to the log-out clean-up functionality?
Q3: Are there issues related to displaying the correct scene headers for any given scene?
Q4: Are there issues related to the question phase of the Game-play scene?
Q5: Are there issues with the information (scores and statistics) in the Game-play scene after playing a game?
Q6: Does the History scene ever show outdated information?
Q7: Are there issues related to stored messages in the Messages scene?
Q8: Are there issues related to updating user information?
Q9: Are there issues with navigating to the correct tab in the Settings scene?

\subsection{Applying MBT on QuizUp}
Modeling was driven by the documentation of the core features of the application, the associated rules, and the derived testing questions. The model was then used to generate abstract test cases, which were then automatically translated into a set of concrete test cases that were executed on QuizUp's Android client. For this study, we used the \emph{Appium} UI test automation tool\footnote{http://appium.io} because 1) it is cross-platform and therefore can be used for both Android and iOS test automation, 2) it is open-source, 3) it is well documented and 4) has a very active community. Appium extends the well-known Webdriver JSON wire protocol specified by Selenium\footnote{http://docs.seleniumhq.org/projects/webdriver}. Tests can be written in any language that has a WebDriver library. Languages such as Ruby, Python, Java, JavaScript, PHP, and C$\#$ all include a WebDriver library. For this study we chose to use Appium's Python implementation of the Webdriver library.
Appium is a web server that exposes a REST API. It receives connections from a client, listens for commands, executes those commands on a mobile device, and responds with an HTTP response representing the result of the command execution. Appium uses the Android UIAutomator\footnote{http://developer.android.com/tools/help/uiautomator} and the iOS UIAutomation tools\footnote{https://developer.apple.com/library/ios/documentation/DeveloperTools/Reference/UIAutomationRef} from the Android and iOS SDK's to inject events and perform UI element inspection. Using the Appium library, UI element objects can be retrieved by UI inspection. Gestures, clicks and keyboard inputs are examples of methods that can then be applied to the retrieved UI element objects.

\subsubsection{The modeling goal}
The primary modeling goal was to design the model so that the derived test cases would be realistic albeit unusual while answering the testing questions and determining whether QuizUp is behaviorally consistent with the requirements. A second modeling goal was to design the model in such way that the derived tests from the model could run on QuizUp's production server. The production server hosts the live version of the application that global users can access. That means that we, as testers, do not have full control of the data space. New users can emerge, the QuizUp team can update the list of available topics, and messages can arise randomly from real users. Thus, we would have to 
design the model in such way that the implemented test code in the mapping table would not be dependent on specific data in the application, but rather implemented to identify types of data and select elements dynamically. 

The third modeling goal was to handle users with different levels of maturity. That is, the derived tests should not be dependent on a particular test user or his status in the game. An arbitrary test user could therefore vary from being a new user who just finished signing up to being an advanced user who has played hundreds of games. 


\subsubsection{The QuizUp model as hierarchical EFSMs}
We modeled the QuizUp application as a collection of EFSMs that were structured in a hierarchical fashion using five layers to manage complexity. It is worth mentioning that some scenes in the application can be accessed as sub-scenes from many different scenes. For example, the Game-play scene can be accessed from scenes such as the Home scene, Topics scene and more. Thus, the layer depth of the Game-play scene itself can vary depending on which scene provides access to it. 

\begin{figure}[t]
    \centering
    \includegraphics[width=3.5in]{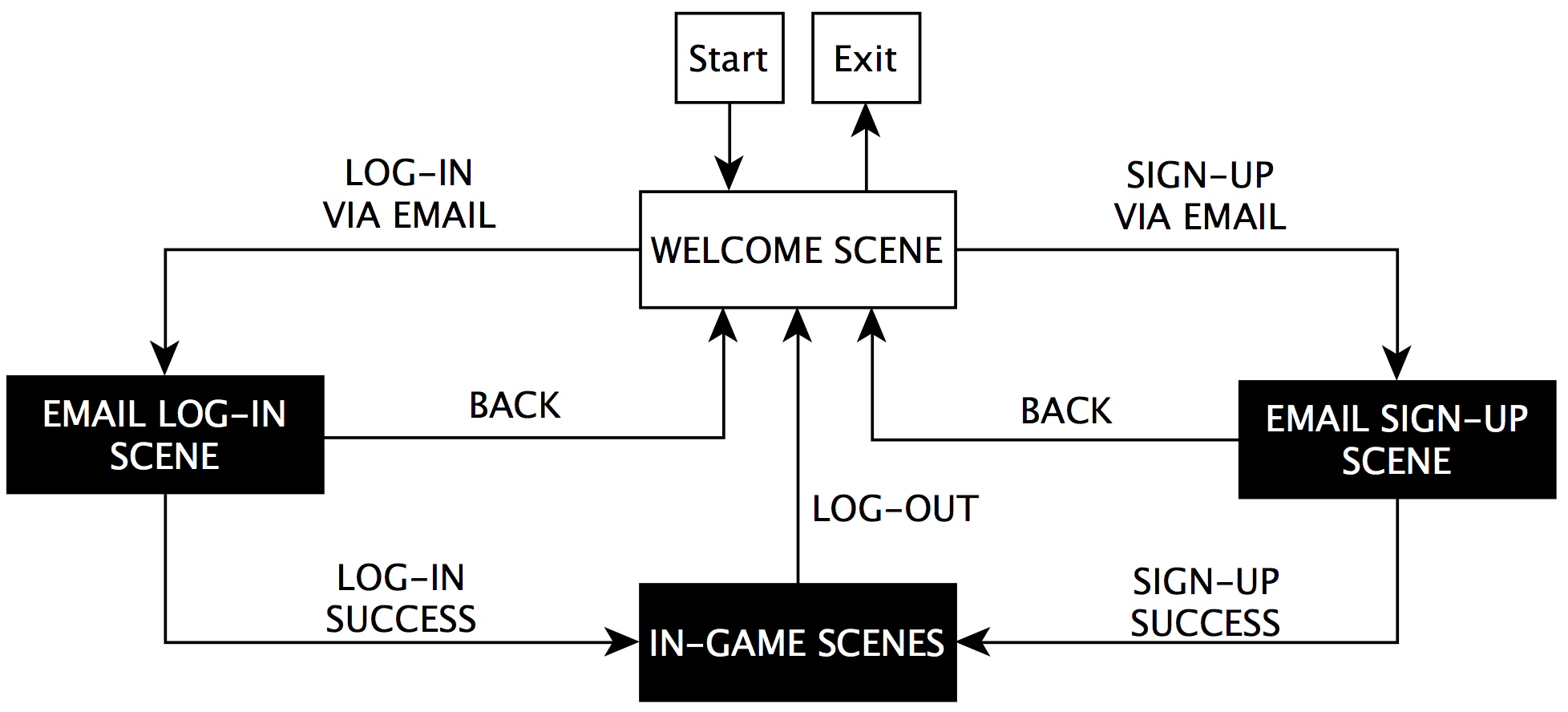}
    \caption{Highest layer of the QuizUp model.}
    \label{fig:quizup_efsm_1st_layer}
\end{figure}

The highest layer (see Figure~\ref{fig:quizup_efsm_1st_layer}) of the model is concentrated on using the Email Log-in or Email Sign-up scenes before entering the application's In-game scenes, where each state for these scenes serves as an entry state to the second layer of the model. The design, thus, explicitly ensures that the `dummy' user (the test case) has logged in prior to using any of the core in-game features. Unusual but valid sequences, such as repeatedly entering and exiting these scenes, are possible outcomes from a random traversal of this model. 

As a result of how we designed the highest layer of the model, the second layer of the model is both concentrated on the step-by-step actions of logging in or signing up, as well as accessing the core features of the in-game scenes. The model is designed in such way that it explores realistic, as well as unusual, sequences for inputting the users credentials (email and password) in order to access the application. Prior to inputting the credentials, a "dummy" test user is retrieved from a service called QTDS (QuizUp Test Data Service), which was implemented specifically for this study. The service is currently very minimal and can be viewed as a partial mock-up of QuizUp's databases. It stores the credentials and key information (e.g. name, title, country) of different test users in a JSON file.
Additionally, the user maturity level is logged by using a helper function, since some in-game scenes (e.g. Messages scene) display different information depending on the maturity of the user. 

The design of the model permits robustness testing (testing of invalid sequences). Based on the documentation, we categorized the types of emails and passwords that a user can input. There are three types of emails to input; a valid email, an email for a non-existing user (invalid), and a badly formed email (invalid). There are two types of passwords to input: a valid password and an invalid password. The model permits any combinations for inputting these different types of emails and passwords. Each transition for these email and password types has an associated helper function that helps us identify which combination was traversed. We are then able to embed the appropriate test oracle for validating the traversed combination. The `validEmailLogin' guard only returns true if both a valid email and a valid password are provided as input. Otherwise, the guard returns false. The `invalidEmailLogin' guard then allows us to understand which type of error message we should expect to be displayed depending on the type of invalidity of the input combination.

\begin{figure}[t]
    \centering
    \includegraphics[width=3.25in]{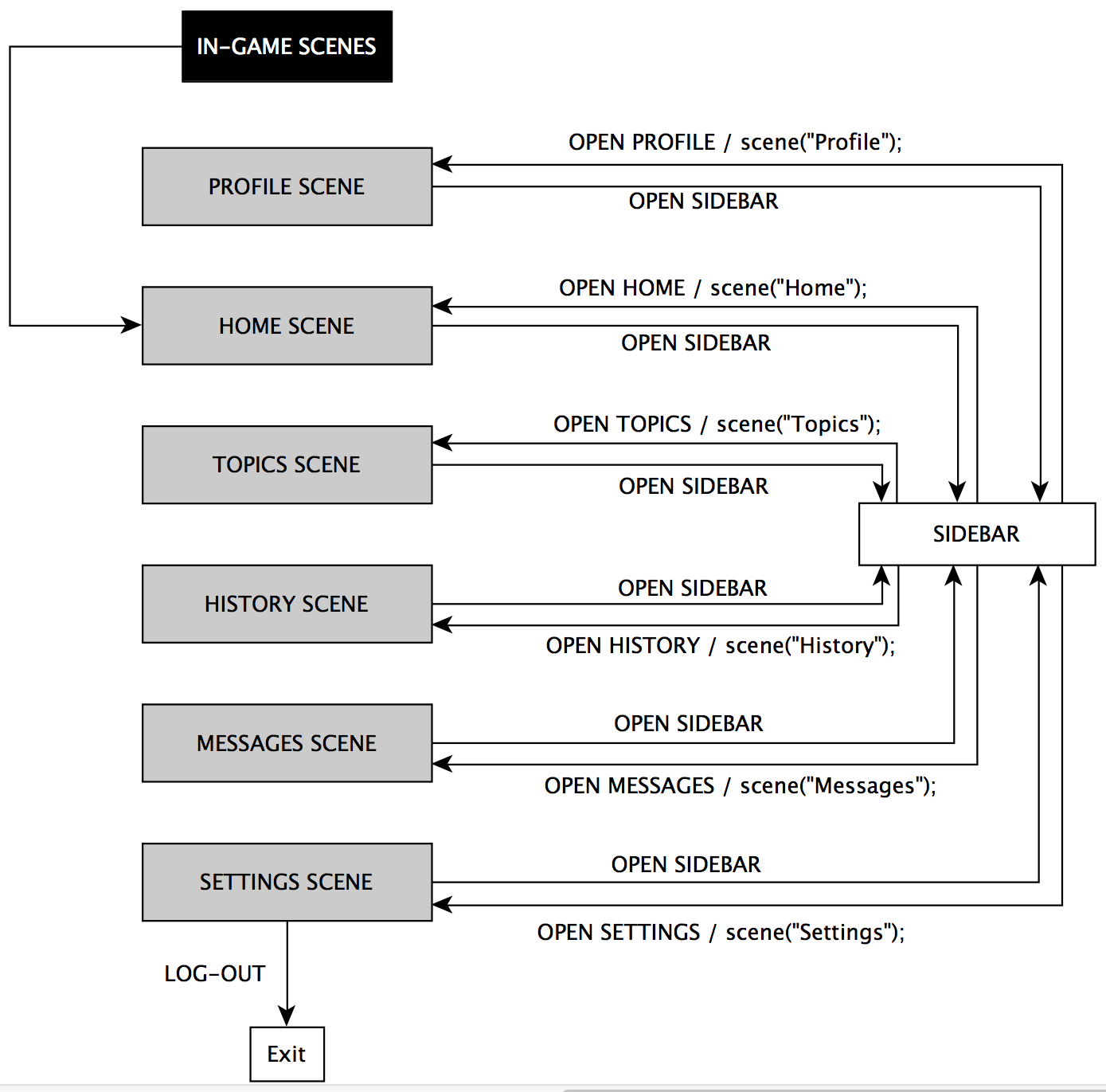}
    \caption{Sub-model for the In-game scenes in the second layer of the model.}
    \label{fig:quizup_efsm_ingame}
\end{figure}

The second-layer model for the Email Sign-up scene has a similar structure and emphasis as the model for the Email Log-in scene. The second-layer model for the in-game scenes of the application is shown in Figure~\ref{fig:quizup_efsm_ingame}. The model includes entry states for the \emph{Profile}, \emph{Home}, \emph{Topics}, \emph{History}, \emph{Messages} and \emph{Settings} scenes, as well as a state for the application's sidebar. Each of the entry states has a sub-model (in the third layer of the model). Therefore, the model is designed as an intermediate layer for navigating between different in-game scenes using the sidebar. When the `dummy' user is led to a new scene during the traversal, the helper function `Scene' is used to add that particular scene to an array which contains the history of traversed scenes for a particular traversal. The function allows us to later embed assertions that rely on the history into our model. As described above, a user can log out from the Settings scene. Thus, a transition for logging out is available from the Settings scene and will bring the user back to the Welcome scene.


    ~ 
\subsubsection{Generation of abstract test cases}
A model traversal starts at the start state in the highest layer of the model and stops when a given stopping criterion is met. We used \emph{yEd}\footnote{http://www.yworks.com}, to create the models in this study and used \emph{Graphwalker} to traverse the models\footnote{http://www.graphwalker.com}. Graphwalker is an open-source Java library that generates test sequences from FSMs and EFSMs that are, for instance, created using yEd models and stored in Graphml format. Graphwalker offers stopping criteria such as \emph{state coverage} and \emph{transition coverage}. We used \emph{Graphwalker} to generate $100$ abstract test cases with built-in assertions for the core features using the random-path algorithm covering $100\%$ of states and transitions. The number 100 is a balance between the costs and benefits of generating and running many test cases. Generating a large number of test cases does not cost anything, since there is no effort or significant time associated with test case generation. Running a test case does not cost anything either in terms of manual effort. However, it takes a relatively long time to execute a test case. For practical purposes it turns out that 100 test cases (in this case) were a good number because the tester could start the test execution, leave for a couple of hours or the night, and then come back and analyze the results.

\subsubsection{Mapping abstract labels to concrete code fragments}
We used Appium's Python library to communicate with the QuizUp application. The code fragments for each label in the QuizUp model were manually inserted into the mapping table. We wrote a script that automatically tracks whether new states or transitions were added to the model, and if so, it generates a template code fragment for new labels.

We learned from the GMSEC case study that non-executable mapping tables can be messy and that maintaining a single mapping table becomes challenging when the model size increases~\cite{DBLP:journals/isse/GudmundssonSGLW15}. Thus, we decided to implement a class structure which would allow us to have a separate mapping table for each scene. Each scene class inherits a scene base class. The base class includes generic code fragments for actions accessible from all scenes (e.g. pressing the Android back button). Therefore, it was unnecessary to implement code fragments for such actions in each scene class. 

\subsubsection{Executing the concrete test cases}
For this study, we implemented a small command-line tool called \emph{Kelevra} intended for MBT projects. The goal for the tool is that both inexperienced and experienced individuals in MBT can use it as a unified interface to trigger the various phases of MBT. Using Kelevra's \emph{instantiate} command we were able to automatically create concrete test cases from the set of $100$ abstract test cases. We then used Kelevra's \emph{run} command with Appium configured as an argument to execute the test suite on two different Genymotion emulators. The Google Nexus 5 and Samsung Galaxy S4 devices were the chosen devices to emulate.

\begin{table}[h!]
\renewcommand{\arraystretch}{1.3}
\caption{Questions that detected issues.}
\label{tab:quizup_analysis_questions_answers}
\centering
\begin{tabular}{| p{0.5cm} | p{9cm} | p{0.85cm} | p{0.85cm} | p{0.85cm} |}
\hline
ID & Testing Question & Ans. & UI & DATA \\
\hline
Q3 & Are there issues related to displaying the correct scene headers for any scene? & \textbf{Yes} & No & \textbf{Yes} \\
\hline
Q8 & Are there issues related to updating user information? & \textbf{Yes} & No & \textbf{Yes} \\
\hline
\end{tabular}
\end{table}

\subsubsection{Analyzing the results and effort}
See Table~\ref{tab:quizup_analysis_questions_answers} for a summary of the results from executing the concrete test cases. In Column $3$, an answer to each testing question is provided. Column $4$ is devoted to issues that can be traced back to badly formed or missing UI elements. Column $5$ deals with issues that can be traced back to incorrect data. The value in a particular cell indicates whether we detected issues related to a particular testing question. 

\begin{ui_issue}
We detected an issue in the key scene list in QuizUp's sidebar in relation to Testing Question 8. The issue, which was not known to the QuizUp team despite their testing effort, originated in the Settings scene (Profile tab) when the `dummy' user updated his name. The name should thereafter be updated in all scenes. However,  this was not the case and therefore the test case failed.
\end{ui_issue}
The EFSM model for QuizUp consists of $125$ states, $192$ transitions, $87$ requests (method calls), and $74$ assertions. The reason for the significant difference between the number of transitions and the number of states is that actions such as opening the sidebar and navigating back using the Android back button were accessible from a majority of the states. The high number of assertions is due to the fact that we were able to embed appropriate test oracles using EFSMs. 


See Table~\ref{tab:quizup_setup_effort} for effort spent before modeling and testing the QuizUp application. We had learned MBT and the Graphwalker tool through the GMSEC study \cite{DBLP:journals/isse/GudmundssonSGLW15}, therefore, no such effort was required for this study. However, applying MBT on a mobile application was new to us and we had to familiarize ourselves with the Appium test driver. We also evaluated other options, such as using the Android UIAutomator directly. We also spent time setting up the Genymotion emulators. Finally, we had to implement a test driver for Appium that was plugged into Kelevra.

\begin{table}[h!]
\renewcommand{\arraystretch}{1.3}
\caption{MBT effort}
\label{tab:quizup_setup_effort}
\centering
\begin{tabular}{| p{11cm} | p{1.25cm} |}
\hline
Task & Effort \\
\hline
Understanding test automation tools, setting up Genymotion emulators & 2 weeks\\
\hline
Implementing the Appium test driver customized for QuizUp & 1 week \\
\hline
Designing the EFSM model and mapping tables & 5 weeks \\
\hline
Executing test cases and analyzing the results & 1 week \\
\hline
\end{tabular}
\end{table}

    

We then spent time understanding QuizUp, incrementally designing the QuizUp EFSM model, and maintained the mapping table as states and transitions were added. During those five weeks we also implemented the test oracles for the EFSM model. There were $15$ guards and helper functions ($112$ source lines of Java-like code that is executed during model traversal) implemented for the QuizUp model. After we finished the model and mapping table, a week was spent on executing the generated test cases and analyzing the results. Thus, it took $9$ weeks for a programmer who was familiar with MBT (but not with mobile applications) to create a significant EFSM model, generate test cases, instantiate them, execute them on two different emulators, and analyze detected defects.

%% file: chapters/04_discussions.tex
\section{Discussion and future work}
Despite the fact that QuizUp is a mobile system, it has certain similarities with non-mobile system we previously tested allowing us to use the same MBT approach again. The key similarity is that QuizUp is a state-based system, where the user interaction can be described as sequences of events in the form of stimuli and responses. This stimuli-response pattern allowed us to model the app as a state machine, which is the key component of the MBT approach used in this study. Most reactive systems can be described in the same way. 
However, some reactive systems do not provide a detailed enough response to all stimuli, making it difficult to determine their state.

We will now compare the QuizUp results to our most recent study on GMSEC~\cite{DBLP:journals/isse/GudmundssonSGLW15}.
At the time of the QuizUp study, we had gained experience in MBT which meant that we were much more efficient in setting up the QuizUp study ($3$ weeks compared to $7$ weeks in the previous study). Based on the results from the previous study, we were also able to immediately determine that EFSMs would be the most appropriate model representation for QuizUp without first experimenting with FSMs. When comparing the model size (i.e. numbers of states and transitions) for GMSEC and QuizUp, it is clear that the QuizUp model is much larger. The GMSEC model consisted of $49$ states and $62$ transitions while the QuizUp model consisted of $125$ states and $192$ transitions. The reason for the difference is twofold. Firstly, we tested more features for QuizUp than for GMSEC and, secondly, in QuizUp, we were able to retrieve more detailed information about the state of the system at any given time. In QuizUp, for example, error messages (text strings displayed in the GUI) were separate for different scenarios or sequences. On the other hand, for GMSEC, we only used simple API return codes to assert whether an action should be successful or not.

There are $8$ requests and $15$ assertions for the GMSEC model, but $87$ requests and $74$ assertions for the QuizUp model. The reason for the large difference is that a state in GMSEC reflects the expected behavior of the system after a particular API method call, whereas a state in QuizUp is comprised of dozens of UI elements that formed a view or a scene in the application. Therefore, for some states in QuizUp, we had to validate multiple UI elements, and data, in order to be confident that we were in fact in a particular state.

An idea based on the study is to create a protocol for modeling mobile applications. Since mobile applications are built using platforms such as Android and iOS, we could provide a standard protocol for modeling common patterns and UI behaviors. The protocol would probably only cover a small subset of possible patterns, but it would be convenient to use the suggested models provided by the protocol as templates when modeling. However, we did not construct a protocol for modeling mobile applications because the projected effort was out of scope for this study.

In addition to a mobile model protocol we discussed adding a simple textual language to describe the present UI elements and patterns in a particular view or scene in a mobile application. For example, `Click Button Ok' could be a statement that would be compiled, or interpreted, to a transition in a model. This could especially help inexperienced individuals in MBT with little modeling experience. However, since we did not construct a mobile model protocol, we decided not to design and implement a textual language.

A couple of notable studies of mobile testing can be found in the literature.
In \cite{6786194} Amalfitano et al.~focus on testing the GUI of a mobile app to find bugs. Since mobile apps are extremely state sensitive, they use test-case generation based on state machines (as opposed to the `stateless' event-flow graphs they employed in their earlier work). They also develop techniques to handle security in mobile apps. The MobiGUITAR framework is implemented as a tool chain that executes on Android. MobiGUITAR uses ripping (an automatic method for creating a state machine model of the GUI of the app), test generation (from the model and test adequacy criteria) and test execution (of tests in the JUnit format in the current implementation). They applied the tool to test four apps from Google Play: Aard Dictionary, Tomdroid, Book Catalogue and WordPress Revision 394. Their testing revealed ten bugs from 7,711 test cases in total. Their conclusion is that the combination of model learning (their ripping phase) and model-based testing is promising for achieving better fault detection in Android app testing. The main difference is that Amalfitano et al.~rip the GUI from the app automatically and create a model from it, while our model is based on the requirements and other artifacts that describe the app. As far as we understand, in the work by Amalfitano et al. the test cases do not have oracles and determine failure based on crashes, while our model has oracles for every request-response pair.

In \cite{deClevaFarto20153}  de Cleva Farto and Endo address similar research questions to ours and also focus on apps for Google Android. They applied MBT to test the AddressBook app modelling the behaviour of the SUT as an Event Sequence Graph (ESG). The model was created manually as in our study. They develop abstract test cases from the ESG models and then make them concrete and execute them using the Robotium platform. The research questions addressed in that paper are very similar to ours and so are their conclusions. The main difference is that  de Cleva Farto and Endo let three groups conduct MBT as an experimental study for a short time (less than an hour), whereas our study spanned three months of work in order to conduct in-depth testing of a commercial app. It is also unclear whether and how oracles are modeled in their approach, while oracles are modeled as an integral part of our methodology.


%% file: chapters/05_conclusions.tex
\section{Conclusions}\label{sec:conclusions}
We presented an empirical study where MBT was applied to the Android client of QuizUp through its GUI. The main goal of the study was to examine whether MBT could be used, in an effective and efficient way, to test mobile systems using the same approach that has been applied to other types of systems, and whether this would be feasible with reasonable effort. 

The study shows that we were able to use an EFSM-based MBT approach to test a mobile system. 
Although QuizUp is indeed different from other systems we have tested (e.g. QuizUp has a graphical user interface, GMSEC does not), there are certain similarities that allowed us to use the same MBT approach. The most notable similarity is that both systems are reactive, state-based systems where the user interaction can be described as sequences of events. 
We found that maintaining a single behavioral model was key in order to test the QuizUp app in an efficient way. This was demonstrated, for example, by the fact that the test cases were able to detect non-trivial issues in a system that was already well-tested. Regarding the effort, as a comparison, much of the effort in the GMSEC study was devoted to learning and applying MBT for the first time, whereas our gained MBT experience allowed us to set up faster and perform a more extensive modeling effort for QuizUp. The effort data and the detected defects show that MBT started paying off as soon as we had applied the process for the first time. We also found that MBT provides a systematic way to test mobile systems and, even though there are still manual steps involved, it is possible to achieve a high degree of automation with reasonable effort for someone who has no or little previous experience with MBT. A possible extension to our work would be to minimize the manual steps even more. Constructing a protocol for modeling mobile applications, for example, would be beneficial to standardize the modeling effort. Another option would be to implement a language to describe a SUT and its possible actions. For mobile applications, we would describe the UI elements and patterns in a particular view or scene under test. The constructed textual description could then be translated into a model representation such as EFSM.